\documentclass[10pt, journal, compsoc, onecolumn]{IEEEtran}

\usepackage[utf8]{inputenc}
\usepackage[T1]{fontenc}
\usepackage{graphicx}
\usepackage{url}
\usepackage{hyperref}
\usepackage{booktabs}
\usepackage{array}
\usepackage{tabularx}
\usepackage{longtable}
\usepackage{soul} % for \hl (highlight/mark)
\usepackage{textcomp}

\hypersetup{
    colorlinks=true,
    linkcolor=blue,
    citecolor=blue,
    urlcolor=blue
}

\begin{document}

\title{Linux and High-Performance Computing}

\author{David~A.~Bader\\
New Jersey Institute of Technology\\
Newark, NJ 07102, USA}

\IEEEtitleabstractindextext{%
\begin{abstract}
In the 1980s, high-performance computing (HPC) became another tool for research in the open (non-defense) science and engineering research communities. However, HPC came with a high price tag; the first Cray-2 machines, released in 1985, cost between \$12 million and \$17 million, according to the Computer History Museum, and were largely available only at government research labs or through national supercomputing centers. In the 1990s, with demand for HPC increasing due to vast datasets, more complex modeling, and the growing computational needs of scientific applications, researchers began experimenting with building HPC machines from clusters of servers running the Linux operating system. By the late 1990s, two approaches to Linux-based parallel computing had emerged: the personal computer cluster methodology that became known as Beowulf and the Roadrunner architecture aimed at a more cost-effective supercomputer. While Beowulf attracted attention because of its low cost and thereby greater accessibility, Roadrunner took a different approach. While still affordable compared to vector processors and other commercially available supercomputers, Roadrunner integrated its commodity components with specialized networking technology. Furthermore, these systems initially served different purposes. While Beowulf focused on providing affordable parallel workstations for individual researchers at NASA, Roadrunner set out to provide a multi-user system that could compete with the commercial supercomputers that dominated the market at the time. This paper analyzes the technical decisions, performance implications, and long-term influence of both approaches. Through this analysis, we can start to judge the impact of both Roadrunner and Beowulf on the development of Linux-based supercomputers.
\end{abstract}
}

\maketitle

\IEEEdisplaynontitleabstractindextext

\section{Introduction}

The Beowulf and Roadrunner projects represent two distinct philosophies for translating these theoretical advances into working systems, and their divergent approaches illuminate enduring tensions in parallel system design between accessibility and performance, standardization and specialization, that remain relevant to contemporary challenges in exascale computing and beyond.

This historical analysis contributes to the anniversary celebration by documenting a pivotal moment when academic innovation and open-source collaboration fundamentally altered an industry previously controlled by a handful of vendors. By 2017, Linux-based systems achieved complete dominance of the TOP500 supercomputer list---a remarkable outcome that vindicated the commodity cluster approach but obscured the distinct contributions of its early pioneers. Understanding which architectural decisions proved prescient and which represented evolutionary dead ends provides lessons for today's researchers navigating similar inflection points in heterogeneous computing, AI accelerators, and quantum-classical hybrid systems. The comparison presented here demonstrates that effective system design requires balancing cost accessibility with performance requirements---a lesson as pertinent to the next 35 years of parallel and distributed computing as it was to the transformative era this paper examines.

Computer systems can tackle larger problems by distributing work across multiple machines connected through a network. When these systems are built using standard, off-the-shelf hardware components rather than specialized equipment, this approach is called commodity cluster computing. This cost-effective method leverages inexpensive, commercially available servers and networking gear to create powerful distributed computing systems that can rival the performance of expensive supercomputers. This work builds upon and significantly expands the historical account presented in the author's IEEE Annals of the History of Computing manuscript ``Linux and Supercomputing: How My Passion for Building COTS Systems Led to an HPC Revolution'' \cite{bader2021linux}. COTS stands for Commercial Off-the-Shelf, referring to ready-made products or software that can be purchased and used immediately, rather than being custom-developed for specific requirements. While that anecdotal piece provided a first-person narrative of developing the Roadrunner Linux supercomputer, this present article conducts a comprehensive comparative analysis of the two foundational approaches to commodity cluster computing: Beowulf and Roadrunner, the latter the work of the present author. This expanded study provides detailed technical analysis of both architectures, systematic comparison through structured tables, expert testimonials from scientific users, and thorough examination of the divergent design philosophies that shaped these seminal systems. Additionally, this paper incorporates substantial new historical research into the Beowulf project's development, technical limitations, and positioning within the broader HPC ecosystem---elements that were not addressed in the prior personal account. The result is an assessment of how these two distinct approaches influenced the evolution of Linux-based supercomputing and established the architectural foundations for modern high-performance computing.

\subsection*{From Big and Bulky to COTS and Open Source}

Vector machines, a type of computer architecture designed for high-speed processing of numerical calculations using Single Instruction, Multiple Data (SIMD) vector processors, had dominated supercomputing since the introduction of the Cray-1 in 1976 \cite{chm_cray1}. The Cray-1 was the first commercially successful supercomputer using vector instructions, and it made the technology mainstream in the supercomputing world. Other systems were developed in the 1980s, including massively parallel multiprocessor systems such as Thinking Machines' CM-5 Connection Machine, launched at the Massachusetts Institute of Technology by W.\ Daniel Hillis and Lewis W.\ Tucker \cite{hillis1993cm5}. Cray introduced a new system called X-MP in 1982, featuring the company's first shared-memory parallel vector processor. An update to that system was the Cray Y-MP, which featured more memory and faster performance \cite{cray_ymp}. The Cray-2 debuted in 1985. They soon replaced the MP systems as the world's fastest machines and were known for their distinctive design featuring total immersion cooling, using a special liquid to cool the densely packed circuit boards. These systems were big: The Cray-1 occupied 2.7m $\times$ 2m of floor space and contained 60 miles of wires \cite{chm_cray1}. They were expensive: in 1976 that same Cray-1 sold for as much as \$10 million. And, unless you worked in national defense, were part of a research team at a large government or academic lab, or with a major industrial user, these machines were out of reach. They ran on proprietary hardware and software, and nothing was compatible with other systems. Moreover, the software itself became harder to create over time. ``The architectures of these systems pose major software problems that the computing industry is ill-equipped to handle, especially for special-purpose systems with limited markets,'' stated the 1982 foundational ``Lax Report'' \cite{lax1982} that led directly to NSF establishing supercomputer centers in 1985 and represented early government recognition of the growing software challenges in supercomputing. Eugene Brooks of Lawrence Livermore National Laboratory predicted that vector processors would not be able to keep up with the new generation of microprocessors using scalar instructions as far back as 1990, when he delivered a talk at Supercomputing 1990 called ``Attack of the Killer Micros'' \cite{brooks1990}. Clearly, improvements in the field were needed, and they came in the form of clustered high-end servers running Linux, a formula for supercomputers that still dominates supercomputing today.

The ASCI Red supercomputer \cite{mattson1996}, deployed at Sandia National Laboratories in 1996--1997, represented the pinnacle of traditional vendor-built massively parallel processing at the time, achieving 1.34 TFLOPS with 9,216 Intel Pentium Pro processors organized in a $38 \times 32 \times 2$ mesh topology. While ASCI Red shared with both Beowulf and Roadrunner the use of commodity commercial off-the-shelf (CCOTS) microprocessors, the similarities largely ended there. ASCI Red relied on proprietary operating systems---the Intel Paragon OS (a distributed UNIX) for service and I/O nodes, and Cougar, a lightweight kernel derived from Sandia's Puma, for compute nodes---rather than Linux. Its custom Interconnection Facility (ICF), built with proprietary VLSI components (the Mesh Routing Component and Network Interface Component), delivered 800 MB/sec bidirectional bandwidth between nodes, far exceeding what commodity Ethernet could provide. However, unlike the early Linux-based approaches, ASCI Red required a massive budget (approximately \$46 million) and a large engineering team from Intel, demonstrating that while commodity processors had become viable for supercomputing, the full system integration---operating system, interconnect, and software stack---remained within the domain of traditional supercomputer vendors. The Linux-based systems discussed in this paper demonstrated that comparable architectural decisions could be implemented at far lower cost using open-source software and, in Roadrunner's case, commercially available high-performance networking.

My journey toward COTS supercomputing began in junior high school in 1981, when I discovered an article about a parallel computing system designed for image processing and pattern recognition \cite{siegel1981}. The concept immediately captivated me---I knew I had to build my own parallel computer. Eight years later, as an undergraduate student at Lehigh University in 1989, serendipity provided the perfect opportunity. While exploring the university's resources, I stumbled upon several donated Commodore Amiga 1000 personal computers gathering dust in a forgotten closet. These machines, combined with my growing knowledge of parallel processing, gave me all the ingredients I needed to construct my first parallel computer and explore new applications that required more computational power. The Amiga cluster nodes ran Commodore's AmigaOS 1.0 operating system, and programming was done through node programs that could send messages over a network to each other. Tools such as job launchers were not provided and needed to be developed.

A year later in 1990, I designed parallel divide and conquer algorithms for combinatorial problems such as sorting and searching on a 128-processor nCUBE hypercube parallel computer donated by AT\&T Bell Laboratories. Programming the nCUBE was very different and required the user to write separate host and node programs. The nCUBE host processor ran the AXIS operating system similar with UNIX, ``but was lacking many of the most useful utility programs that UNIX users are accustomed to having'' \cite{tolle1988}. These early experiences taught me that the development of powerful parallel machines required a simultaneous development of scalable, high-performance algorithms and services. Otherwise, application developers would be forced to develop algorithms and basic tools from scratch every time vendors introduced newer, faster hardware platforms.

At the University of California, Berkeley, a pioneering distributed computing research project was launched in the mid-1990s called Network of Workstations (NOW), led by David Culler \cite{anderson1995}. Concurrently, Miron Livny, head of the University of Wisconsin's Condor project, aimed to make a building full of desktop computers act as a single large computer \cite{litzkow1988}. These projects leveraged high-bandwidth, switch-based local area networks with a custom low latency network interface and global layer operating system. NASA began the Beowulf project in 1994, following the ``Pile-of-PCs'' methodology to build a parallel workstation from a cluster of personal computers (PCs). In their 1995 paper, ``Beowulf: A Parallel Workstation for Scientific Computing'' \cite{sterling1995icpp}, Beowulf developers Thomas Sterling \emph{et al.}\ describe ``the Beowulf parallel workstation [as] a single user multiple computer with direct access keyboard and monitor.'' The term ``multiple computer'' refers to combining multiple PCs into a unified system to increase processing power and computational capabilities.

For over a decade, the computing world accepted a compelling origin story: Beowulf clusters were named after the epic hero who challenged formidable opponents, symbolizing how commodity hardware could take on expensive supercomputers. This narrative, told consistently by creators Thomas Sterling and Donald Becker from 1994 to 2004, fit perfectly with their David-versus-Goliath approach to parallel computing. But in NASA's 2020 Spinoff publication, Sterling finally revealed the truth. The naming was completely accidental and occurred under pressure. When a NASA Goddard Space Flight Center (GSFC) program administrator called demanding an immediate project name in 1994, Sterling was ``helplessly looking around for any inspiration'' in his office. His mother had majored in Old English, leaving him with a copy of the Beowulf epic. In desperation, he told the administrator: ``Oh hell, just call it Beowulf. Nobody will ever hear of it anyway'' \cite{nasa2020spinoff}. Sterling admits the heroic justification was ``invented in hindsight'' for public relations purposes. The fabricated narrative served the project well, providing a memorable and meaningful explanation that resonated with the cluster computing community's understanding of their battle against established supercomputer manufacturers.

Beowulf clusters targeted individual researchers who needed affordable parallel computing workstations. Beowulf, according to NASA's James R.\ Fischer, was created to provide NASA staff with gigaflops workstations that would allow them to use and share software over various platforms \cite{fischer2014}. The frustration of dealing with difficult-to-use and often incompatible hardware and software motivated the project, as well as the need for better price-to-performance ratios. Its scope was limited to building a prototype scalable workstation for NASA's scientists that could run the same Earth and Space Sciences software (but not as efficiently) as supercomputers of the day. However, the Beowulf project did show that commodity hardware and open-source operating systems and software had a role to play in the world of supercomputing. C.\ Gordon Bell, National Academy of Engineering (NAE) member, Microsoft Emeritus Researcher and founding Assistant Director of the National Science Foundation's Computing and Information Science and Engineering Directorate, and colleague Jim Gray, then manager of Microsoft Research's eScience Group, described the Beowulf technology in the early 2000s as ``Beowulf enabled do-it-yourself cluster computing using commodity microprocessors'' \cite{bell2002}. By that time, Beowulf systems offered a single platform standard that allowed applications to be written and to run on more than one computer. Bell and Gray acknowledged that the Beowulf framework, while cost effective, could not meet all supercomputing needs and ``perform[ed] poorly on applications that require large shared memory'' \cite{bell2002}. They encouraged research into next-generation ``Beowulfs'' that would stimulate cluster understanding and training so they can serve the needs of research centers that depend on high-end supercomputing.

While the Beowulf project zeroed in on price/performance, a different focus motivated me, then a professor of electrical and computer engineering at the University of New Mexico, as I took the concept of using commodity clusters for speed and compatibility at a lower price point but prioritizing performance above all. My alternative Linux-based supercomputing architecture combined Linux, COTS components, and (newly) high-speed, low-latency interconnection networks. Bell, looking back on these efforts, commented that ``Bader was first to design a Linux supercomputer with the speed, performance and services of a large, centralized and general-purpose supercomputer'' \cite{bell2021}. In April 1998, the Roadrunner Phase~1 system used the new Myrinet System Area Network (Myrinet/SAN) \cite{boden1995}, which was about 256 times the total bandwidth available using standard Ethernet on which Beowulf's networks ran. I named the system Roadrunner after the New Mexican state bird, which is the fastest running bird capable of flight. This name combined ``speed'' and ``New Mexico'' together with mental images of the cartoon Roadrunner outsmarting Wiley E.\ Coyote. Roadrunner also included compilers, a job scheduler, and features to enable parallel programming, such as software-based distributed shared memory and Message Passing Interface (MPI). The Roadrunner Phase~2 system became the first Linux supercomputer available for open use by the national science and engineering community through the National Science Foundation's National Technology Grid, entering production in April 1999 \cite{fleck1999}. Within a decade, this architectural approach became the predominant model for COTS supercomputing, which in turn became the dominant model for all supercomputers worldwide.

Bell, reflecting much later on the development of Roadrunner, highlighted the importance of performance focus, noting the ``Bader Roadrunner design could efficiently run the national science community's most demanding supercomputing applications at a fraction of the cost of traditional supercomputers---unlike Beowulf clusters that were used by individuals and were not competitive in performance'' \cite{bell2021}. The two approaches served different purposes and made different contributions to the evolution of supercomputing. This paper highlights these design approaches, with Beowulf focused on mass-market components and low price points and Roadrunner focused on higher performance at the cost of some specialized components.

\subsection*{Comparative Analysis: Beowulf Vs.\ Roadrunner Architectures}

\begin{table*}[!ht]
\centering
\caption{Comparison of Beowulf versus Roadrunner.}
\label{tab:comparison}
\small
\begin{tabularx}{\textwidth}{@{}l X X@{}}
\toprule
\textbf{Feature} & \textbf{Beowulf} & \textbf{Roadrunner} \\
\midrule
Development Target & Low-cost replacement for single-user workstation & Low-cost replacement for traditional supercomputer \\ \hline
Design Philosophy & ``Mass-market commodity off-the-shelf'' (M\textsuperscript{2}COTS) with strict vendor-neutral requirements & Balanced integration of commodity components with specialized high-performance technologies \\ \hline
Development Timeline & NASA project begun in 1994, first system operational in 1995 & Prototype developed in 1998, production operation April 1999 \\ \hline
User Model & Single-user, single-application workstation & Multi-user, multi-application shared resource \\ \hline
Network Technology & Ethernet & Specialized three-network architecture including 1) high-speed, low-latency interconnects for data network, 2) RS-232 serial network for diagnostics, and 3) an Ethernet control network \\ \hline
Node Architecture & Uniprocessor nodes & Multiprocessor nodes \\ \hline
System Management & None & Tools for job scheduling and resource allocation \\ \hline
Target Applications & Earth and Space Science satellite image processing & Scientific applications from astrophysics and cosmology, climate and weather research, physics research, engineering and applied science, materials science, computer science and systems research, graph analytics, and national security \\ \hline
Positioning & Explicitly designed as complementary to---not a replacement for---commercial HPC systems & Direct alternative to commercial supercomputers \\ \hline
Integration Model & ``Just-in-place'' DIY integration by end-users & Vendor integration \\ \hline
Scalability & 10's of processors & 1000's of processors \\ \hline
Historical Influence & Popularized commodity cluster concept & Established architectural template for modern supercomputing \\ \hline
Development Team & Five engineers (Thomas Sterling, Don Becker, John Dorband, Don Jacob, and Jim Fischer) with NASA institutional support & Single individual (David A. Bader) \\
\bottomrule
\end{tabularx}
\end{table*}

\section{The Beowulf Vision of Personal Parallel Computing}

From their inception, Beowulf systems were conceived as enhanced parallel workstations rather than traditional supercomputers. Sterling and his collaborators were explicit about this positioning. In his 1996 \emph{Communications of the ACM} article, Sterling characterized Beowulf as ``an experimental distributed PC system developed to evaluate this new PopC opportunity in single-user environment computing,'' emphasizing that ``the PopC approach as demonstrated by Beowulf does not so much challenge the workstation market as reveal a possible direction for its advance and advantage'' \cite{sterling1996cacm}. The 1995 \emph{HPDC} paper made the design target even more explicit: ``While most distributed computing systems provide general purpose multi-user environments, the Beowulf distributed computing system is specifically designed for single user workloads typical of high end scientific workstation environments'' \cite{sterling1995hpdc}. That paper further noted that ``the very nature of Beowulf as a single-user workstation demanded that the marginal replacement cost of the system be consistent with pricing of contemporary high end workstations''---approximately \$40,000 for the prototype \cite{sterling1995hpdc}. These parallel workstations would give researchers a development environment running the same software stack as production supercomputers, enabling code development and testing at smaller scales before scaling up to larger institutional systems. The Beowulf team acknowledged inherent scalability limitations consistent with this role: ``It is true that Beowulf is limited in the scope of its scaling. Its configuration is only intended to exploit the first order of magnitude in parallelism'' \cite{sterling1995hpdc}. The 1997 \emph{IEEE Aerospace Conference} paper reinforced this positioning, stating that the Pile-of-PC methodology ``is not for everyone. Rather, it is an emerging opportunity in the high performance computing field and complements rather than competes with the HPC industry commercial products'' \cite{ridge1997}. The fundamental philosophy emphasized democratizing parallel computing access rather than directly challenging established supercomputing centers. Roadrunner, by contrast, was designed from inception as a general-purpose, multi-user supercomputer serving a national user community, directly competing with commercial supercomputer offerings rather than complementing them.

\begin{figure}[!t]
\centering
\includegraphics[width=3.0in]{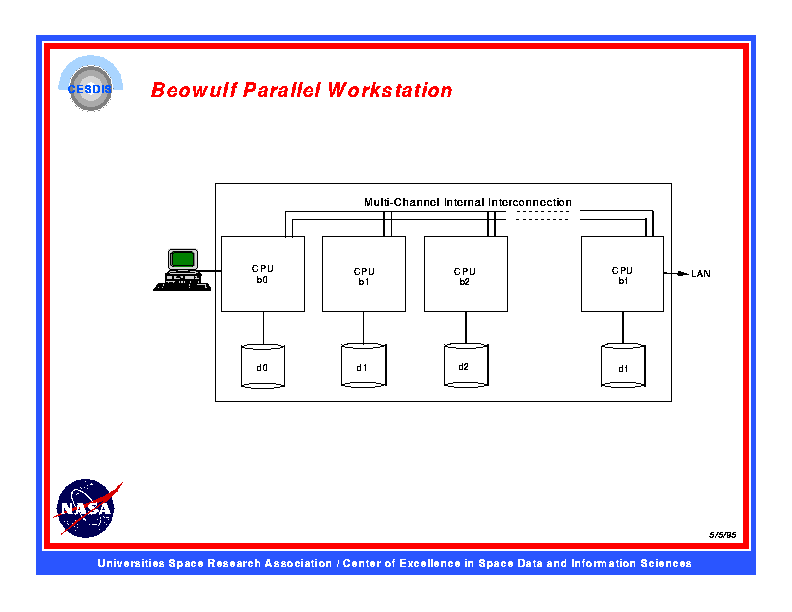}
\caption{NASA Slide of Beowulf Parallel Workstation. Image source: \cite{sterling1995decus}.}
\label{fig:beowulf_slide}
\end{figure}

Beowulf, short for ``Beowulf Parallel Workstation,'' (see Figure~\ref{fig:beowulf_slide}) was envisioned to augment a single user with more compute power using mass market PCs and Ethernet, its creators reported at the \emph{IEEE Aerospace Conference} in 1997 \cite{ridge1997}. Thomas Sterling, one of Beowulf's co-creators, put it this way: ``The Beowulf parallel workstation defines a new operating point in price-performance for single-user computing systems'' \cite{sterling1995icpp}. James Fischer, the project manager for NASA's High-Performance Computing and Communications Earth and Space Science Project at GSFC during the development of Beowulf recollected, ``Looking back to the origins of the Beowulf cluster computing movement in 1993, it is well known that the driving force was NASA's stated need for a gigaflops workstation costing less than \$50,000'' \cite{fischer2014}.

The 1998 NASA workshop assessment confirmed this workstation-centric origin, reporting that Beowulf's charter was ``a single user `gigaflops' science workstation'' and that the primary design consideration was not floating point performance but rather disk access: ``An evaluation of the requirements for a scientific station for NASA showed that disk access capacity and bandwidth was far more important to user response time than was floating point performance'' \cite{sterling1998assessment}. The project's cost target was explicitly set at ``the price of a high-end scientific workstation, assumed to be \$50,000,'' with the goal of assembling a cluster of low-cost PCs with an order of magnitude larger disk capacity and approximately eight times the disk bandwidth of a single workstation \cite{sterling1998assessment}.

\begin{figure}[!t]
\centering
\includegraphics[width=2.5in]{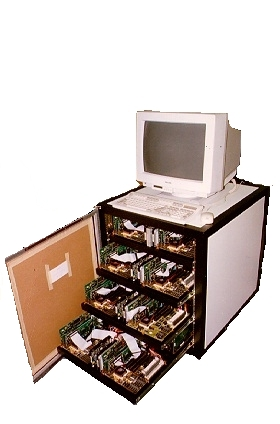}
\caption{Wiglaf, the original Beowulf prototype. Image source: NASA Goddard Space Flight Center.}
\label{fig:wiglaf}
\end{figure}

The original Beowulf prototype, named ``Wiglaf,'' (see Figure~\ref{fig:wiglaf}) was built by Thomas Sterling, Don Becker, John Dorband, and Dan Jacob, at NASA GSFC in late 1994. Wiglaf contained 16 uniprocessor nodes (each with a 66 MHz Intel 80486 and SiS 82471 chipset motherboard), connected by commodity 10BaseT Ethernet. The developers chose the Slackware Linux distribution to run on each node. At the end of 1994, the developers updated Wiglaf's processors, replacing them with 100 MHz Intel 80486 DX4 processors.

\subsection*{Strict Commitment to Commodity Hardware}

The Beowulf design philosophy, in its effort to keep costs down, enforced adherence to mass-market commodity components, rejecting any technology that was not widely available through multiple vendors. This approach was codified as ``M\textsuperscript{2}COTS'' (mass-market commodity off-the-shelf) \cite{katz1998}. In 1997, Michael Warren \emph{et al.}\ published a parallel computing conference paper where they define ``Beowulf-class systems [as] off the shelf M\textsuperscript{2}COTS PCs [that] are interconnected by low cost local area network (LAN) technology running an open source, Unix-like operating system and executing parallel applications programmed with an industry standard message passing model and library'' \cite{warren1997pdpta}. Sterling emphasizes this point in his 1999 How to Build a Beowulf book: ``Beowulfs use only mass-market components and are not subject to the delays and costs associated with custom parts and custom integrations'' \cite{sterling1999howto}. By restricting itself to the most commoditized hardware, Beowulf systems were true to the M\textsuperscript{2}COTS ideal, at the expense of performance potential. Sterling formalized this boundary in the 1998 NASA workshop assessment, stating that ``Beowulf-class systems exclusively employ COTS technology, usually targeted to the mass market where cost benefits of mass production and distribution drive prices down'' \cite{sterling1998assessment}. The assessment acknowledged one exception: networking technology, ``which while COTS, is not mass market due to the relatively small volume.'' However, even for this exception, Sterling imposed a cost ceiling: networking components were included only if vendors could ``provide their product within that cost framework'' \cite{sterling1998assessment}. This formulation reveals a critical distinction: a technology could be commercially available---as Myrinet was---and still fall outside the Beowulf class if it exceeded the cost bounds that defined the M\textsuperscript{2}COTS paradigm. This stance extended to all aspects of system design, including processors, networking equipment, and software components. The use of vendor-neutral components meant Beowulf clusters could not benefit from the performance of specialized hardware, but enjoyed flexibility in system construction. As Sterling explained, ``Basically, you can order most of Beowulf's components from the back pages of Computer Shopper or get them for free over the `Net''' \cite{nasa2020spinoff}.

A \emph{Pile-of-PCs} is the term used today to describe the loose ensemble or cluster of PCs applied in concert to a single problem. Similar in principle to the Berkeley NOW project, it emphasizes mass market commodity components, dedicated processors (rather than scavenging cycles from idle workstations), and a private system area network (SAN), all with the goal of achieving the best system cost/performance ratio. Beowulf added to this the following principles: no custom components, easy replication from multiple vendors, a freely available software base, using freely available distribution computing tools with minimal changes, and returning the design and improvements to the community. The approach exploits components that respond to widely accepted industry standards and benefits from lower prices resulting from heavy competition and mass production. Subsystems provide accepted, standard interfaces such as PCI bus, IDE and SCSI interfaces, and Ethernet communications. This is the Beowulf approach, and one advantage is that no single vendor owns the rights to the product. In a Computer History Museum talk in 2000, Sterling remarks that ``Beowulfs formed a do-it-yourself cluster computing community using commodity microprocessors, local area network Ethernet switches, Linux (and now Windows 2000), and tools that have evolved from the user community. This vendor-neutral platform used the MPI message-based programming model that scales with additional processors, disks, and networking'' \cite{sterling2000chm}. As Steve Elbert noted at the 1\textsuperscript{st} Pentium Pro Cluster Workshop in 1997: ``The difference between Beowulf and other parallel processing systems is that it has no custom hardware or software but consists of standard, off-the-shelf computers, \ldots\ connected by Ethernet and functioning as a single machine on the Linux operating system'' \cite{elbert1997}.

\subsection*{Beowulf Network Choices}

The Beowulf architecture was defined in ways that excluded high-performance interconnects. Sterling's 1999 MIT Press book states that ``A Beowulf-class system is a cluster of mass-market commodity off-the-shelf (M\textsuperscript{2}COTS) PCs interconnected by low cost local area network (LAN) technology running an open source code UNIX-like operating system and executing parallel applications programmed with an industry standard message passing model and library'' \cite{sterling1999howto}, and that systems using specialized HPC networks would not be a Beowulf. The Los Alamos National Laboratory's (LANL) Avalon Project FAQ explained that Myrinet was not used because it would ``double the cost'' of the system and was not considered a mass-market component \cite{lanl_avalon}. The 1998 NASA workshop assessment quantified why Myrinet fell outside Beowulf's design envelope. Sterling and his co-authors established a cost guideline for Beowulf networking: ``A rule of thumb is that the network should cost approximately one quarter of the total system cost. Given that a current node cost is between \$1,600 and \$1,800 per node without communication, the cost for communication per node should be between \$530 and \$600 per node'' \cite{sterling1998assessment}. By contrast, the assessment reported that Myrinet NICs alone cost approximately \$1,400 per port---more than double the entire acceptable per-node networking budget---leading Sterling to conclude that ``a system employing Myrinet can dedicate half of the cost to the communication network; probably not the balance one would choose'' \cite{sterling1998assessment}. Fast Ethernet, at roughly \$225 per node for small systems, comfortably fit within these constraints. The cost differential was not incidental; it was the architectural boundary that defined the Beowulf class.

This narrow definition of ``commodity''---limited to components sold in retail computer stores for desktop systems---had significant architectural consequences. Any network technology designed specifically for HPC, regardless of its commercial availability, fell outside Beowulf's self-imposed boundaries. Roadrunner adopted a broader definition: any component purchasable from a commercial vendor without custom engineering qualified as commodity. Myrinet, though manufactured for the HPC market, was a standard catalog item available to any buyer. This definitional difference was not merely semantic---it determined whether Linux-based clusters could achieve competitive supercomputer performance or would remain limited to embarrassingly parallel workloads.

Sterling's own assessment confirms that commodity processors alone did not qualify a system as Beowulf-class. In the 1998 NASA workshop proceedings, Sterling noted that ASCI Red was ``not a Beowulf-class system'' despite the fact that it ``incorporates the identical chip level technology as that of Beowulf with the exception of the NIC (network interface controller)'' \cite{sterling1998assessment}. This exclusionary principle---that a system sharing Beowulf's commodity processors but employing a different networking architecture fell outside the Beowulf class---applied with equal force to Roadrunner. Roadrunner used commodity Intel processors but integrated Myrinet as its data network, a technology that the same 1998 assessment explicitly treated as incompatible with Beowulf's cost constraints.

However, by limiting systems to commodity Ethernet, Beowulf clusters suffered performance bottlenecks in communication-intensive applications. These bottlenecks restricted the types of computational problems that could be efficiently solved, with limitations most apparent when running tightly-coupled parallel applications requiring frequent inter-node communication. This was particularly problematic since such workloads represented precisely the applications that drove most supercomputing research and development of the era. But for the Beowulf team, cost was the most important consideration as ``the driving force was NASA's stated need for a gigaflops workstation costing less than \$50,000'' \cite{fischer2014}.

In 1994 Sterling recruited Don Becker to NASA for his skill at writing operating system software \cite{nasa2020spinoff}. Because Beowulf's ``processors were too fast for a single Ethernet and Ethernet switches were too expensive'' \cite{sterling1997pileofpcs}. ``To balance the system Don Becker rewrote his Ethernet drivers for Linux and built a `channel bonded' Ethernet where the network traffic was striped across two or more Ethernets'' \cite{sterling1997pileofpcs}. Channel bonding represented an innovative but ultimately temporary solution. In 1995, Fast Ethernet was introduced. At 100 Mbps, Fast Ethernet could transfer data at 10$\times$ the rate of standard Ethernet. At a 1995 conference, Beowulf developers acknowledged that one of the limiting factors for scaling up Beowulf clusters was the network, and they believed Fast Ethernet was the solution \cite{sterling1995decus}. Beowulf developer Michael Warren stated: ``Nothing can beat Fast Ethernet for price/performance at the moment'' \cite{warren1997loki}; and the community rejected Myrinet as ``not commodity'' \cite{elbert1997}. By 1998, industry documentation acknowledged the channel bonding technique was already being supplanted: ``As 100Mbit/s Ethernet and 100 Mbit/s Ethernet switches have become cost effective, the need for channel bonding has gone away (at least for now)'' \cite{redhat1998}.

According to Sterling, the Beowulf project provided improvements over prior personal computer clusters that included: ``Ethernet drivers, channel bonding, advanced topologies, applications, [and] ensemble management tools'' \cite{sterling1997pileofpcs}. The Beowulf developers focused their research efforts on mitigating the relatively long latencies and modest interconnection bandwidth provided by low-cost networking such as Fast Ethernet. ``This is being addressed by software performance tuning, aggregating networks, and rich interconnect topologies,'' the team reported \cite{sterling1996cacm}.

\subsection*{Beowulf Network Topology}

In 1997 Beowulf developers then attempted to overcome Ethernet's limitations by using complex network topologies. These efforts included elaborate multi-network designs, as well as hyperlinked topologies to reduce communication latency through strategic node connections. The most notable example was the ``hypercube plus switch'' topology used in LANL's ``Loki'' cluster \cite{warren1997loki}.

While these complex topologies provided performance improvements for specific communication patterns, they added significant system management. As Fast Ethernet switches became more affordable, this created a new choice for Beowulf designers: continue using complex topologies with software-based routing, or adopt simpler network designs that relied on dedicated hardware switches. As the team reported in 1996: ``The original networking strategy for Beowulf was to use multiple Ethernet networks in parallel, each connecting all the nodes within the system. Both 10 Mbps and the new 100 Mbps Fast Ethernet were employed in separate Beowulf systems. The parallel networks were managed through a technique called channel bonding that uniformly distributed packets among the interconnects in a manner transparent to the user code. \ldots\ It is shown that in many circumstances the more complex topologies perform better, and in some circumstances software routing techniques compare favorably to more expensive hardware switch mechanisms'' \cite{sterling1996cacm}. This research demonstrated that software-based routing in complex topologies could still compete with hardware switching solutions.

\begin{figure}[!t]
\centering
\includegraphics[width=3.5in]{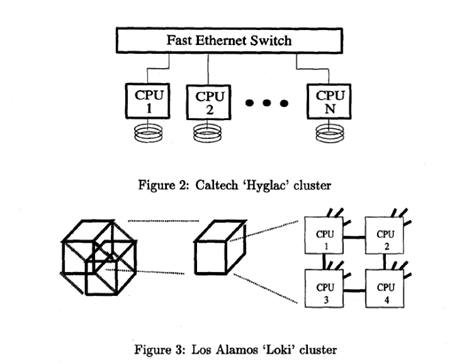}
\caption{Beowulf network topology examples.}
\label{fig:beowulf_topology}
\end{figure}

By this time, major research institutions were deploying 16-node Pentium Pro Beowulf clusters tailored to their specific computational needs. Notable examples included Caltech's ``Hyglac'' cluster and LANL's ``Loki'' machine, identical systems except for their topologies, both optimized for N-body galactic gravitational simulations, as well as a system at NASA GSFC. These installations demonstrated different networking approaches---from Caltech's Fast Ethernet crossbar switch configuration to LANL's hybrid hypercube-plus-switched network design that improved performance by bypassing longer hypercube routes (Figure~\ref{fig:beowulf_topology}) \cite{warren1997loki}.

However, investigations of topologies for Ethernet networks declined when commodity Gigabit Ethernet switches became available in 1999. By 2000, tutorials on building Beowulf clusters advocated for buying these switches: ``Switches have become inexpensive enough that there's not much reason to build your network by using cheap hubs or by connecting the nodes directly in a hypercube network'' \cite{lindheim2000}.

\subsection*{Beowulf Uniprocessor Node Architecture}

Beowulf developers initially avoided symmetric multiprocessing (SMP) nodes in favor of uniprocessor architectures. At the time Roadrunner was being developed, the prevailing Beowulf guidance discouraged multiprocessor configurations. At cluster workshops in 1997, Beowulf advocates argued that ``SMP is a terrible idea'' for three main reasons: research computing centers were already short on memory bandwidth and network bandwidth, and multiple CPUs on a single node were seen as making the problems worse; message passing programs should not require users to also deal with shared memory complexity; and multiple processors sharing memory was seen as making it harder to write a stable and efficient operating system \cite{warren1997loki}. At another 1997 workshop, Beowulf developers stated their view that ``using multiple processors within each node (SMP) is unlikely to be a good idea for many applications'' \cite{sterling1997tutorial}.

The 1998 NASA workshop assessment reinforced this position on technical grounds, reporting that while motherboards with up to four processors per SMP node were available, ``the memory bandwidth is inadequate to support good utilization of these processors except in particularly carefully crafted codes that make particularly favorable usage of the two layers of caches on each processor'' \cite{sterling1998assessment}. The assessment further noted that Linux's SMP support at the time was limited: ``the current implementation does not permit more than one O/S service call to be performed simultaneously so that operating system work does not speed up with additional processors within an SMP node'' \cite{sterling1998assessment}. These constraints reflected genuine technical limitations of the era---limitations that Roadrunner's custom kernel work specifically addressed.

However, the Beowulf community's thinking on this issue evolved as multiprocessor systems became more commodity and cost-effective. By 2001, Sterling's MIT Press volume \emph{Beowulf Cluster Computing with Linux} acknowledged that dual-processor SMP nodes might represent the ``price/performance sweet spot'' for cluster configurations \cite{sterling2001book}. This evolution reflects the Beowulf project's pragmatic response to changing hardware economics rather than rigid adherence to original design principles.

Roadrunner's 1998 adoption of dual-processor nodes thus anticipated a direction that the broader commodity cluster community would eventually embrace. At the time of Roadrunner's development, however, this architectural choice represented a deliberate departure from contemporaneous Beowulf guidance, reflecting my emphasis on computational density and hierarchical parallelism over strict adherence to the uniprocessor model then prevalent in the Pile-of-PCs community.

\subsection*{Beowulf's Contribution to Cluster Approaches}

The Beowulf project's technical innovations focused on administrative improvements. These differences included dedicated (rather than timeshared) compute nodes, isolated network infrastructure, and operating system optimizations. The software contributions focused primarily on system management tools, Ethernet driver modifications, and the channel bonding techniques described previously. The Beowulf team described their differentiating factors thus in 1997:

\begin{quote}
``A Beowulf-class cluster computer is distinguished from a Network of Workstations by several subtle but significant characteristics. First, the nodes in the cluster are dedicated to the cluster. This helps ease the load balancing problem, because the performance of individual nodes are not subject to external factors. Also, since the interconnection network is isolated from the external network, the network load is determined only by the application being run on the cluster. This eases the problems associated with unpredictable latency in NOWs. All the nodes in the cluster are within the administrative jurisdiction of the cluster. For example, the Beowulf software provide a global process ID which enables a mechanism for a process to send signals to a process on another node of the system. This is not allowed on a NOW. Finally, operating system parameters can be tuned to improve performance. For example, a workstation should be tuned to provide the interactive feel (instantaneous responses, short buffers, etc), but in a cluster the node can be tuned to provide better throughput for coarser grain jobs because they are not interacting with users'' \cite{reschke1996}.
\end{quote}

\subsection*{Scope and Positioning within HPC}

Beowulf developers were clear about their design goals and target market, positioning their system as a complement to existing high-performance computing solutions, focusing on providing affordable parallel computing access to individual researchers. This would suit scientists who required parallel computing capabilities but lacked access to traditional supercomputing resources. They wrote: ``The Pile-of-PC methodology is still experimental, and does not match all of the valuable services provided by computer vendors. It is not for everyone. Rather, it is an emerging opportunity in the high-performance computing field and complements rather than competes with the HPC industry commercial products'' \cite{ridge1997}. This positioning was consistent from Beowulf's inception. In their foundational 1995 paper, Sterling \emph{et al.}\ explicitly stated that ``the Beowulf distributed computing system is specifically designed for single user workloads typical of high end scientific workstation environments,'' contrasting it with ``general purpose multiuser environments'' \cite{sterling1995icpp}. Sterling reiterated this vision in 1998, that Beowulf was not aimed at replacing traditional computers. ``Instead, we're trying to augment genuine supercomputers with superior price performance for a lot of real-world problems'' \cite{mcgrath1998}. The 1998 NASA workshop assessment candidly acknowledged these limitations. Sterling and his co-authors reported that while Beowulf-class systems could deliver multi-gigaflops performance at unprecedented price-performance, ``software environments were not fully functional or robust, especially for larger `dreadnought'-scale systems'' \cite{sterling1998assessment}. The assessment noted that ``few users are completely happy in either case at this stage'' and that support for larger multi-user systems ``is still an open question'' \cite{sterling1998assessment}. These acknowledged gaps---in system software maturity, multi-user support, and scalability beyond modest node counts---were precisely the challenges that Roadrunner's design set out to address through its integrated approach to system software, job scheduling, and high-performance networking. Fischer later clarified that Beowulf's goal was ``to complement the large parallel computers they were using'' \cite{ncsu2022}, and that the project arose fundamentally from the need for ``software environments integration'' and software sharing---not from an ambition to challenge supercomputer performance \cite{fischer2014}.

In his 2000 Computer History Museum talk \cite{sterling2000chm} Sterling said Beowulf's contributions were primarily ``the majority of Ethernet drivers'' and ``channel bonding to support multiple simultaneous and many other low level tools for managing clusters of PCs.'' The abstract for his talk says Beowulf systems could ``equal in performance that of much more costly machines'' while providing ``a price-performance advantage of an order-of-magnitude or more.'' Such benefits were, however, restricted to applications that could tolerate the communication bottlenecks inherent in commodity Ethernet networks.

\subsection*{Trade-offs of the DIY Integration Model}

Sterling characterizes the Beowulf approach as providing ``tremendous flexibility'' through its ``just in place'' integration model \cite{sterling1997tutorial}. This approach offered distinct advantages by allowing organizations to customize systems according to their specific needs and budgets. However, this flexibility came with trade-offs that organizations needed to consider when adopting Beowulf clusters, as it would ``more fully engage the talents of the on-site technical staff to enable operation than would a vendor provided turn-key system'' \cite{sterling1997tutorial}.

The DIY integration model required organizations to assume system integration responsibilities typically handled by commercial vendors. This meant that adopting institutions needed to develop or maintain in-house expertise for system assembly, configuration, driver compatibility management, and ongoing maintenance. For organizations with existing technical staff, this represented an opportunity to gain deep system knowledge and maintain full control over their computing environment. However, for institutions without such expertise, these requirements could offset some of the initial cost savings through increased personnel needs or external consulting costs.

This integration model naturally favored organizations with strong technical capabilities and limited the adoption primarily to institutions that could effectively handle the system administration requirements. As a result, Beowulf clusters found their strongest adoption among technically sophisticated users who valued the flexibility and cost control that the approach provided. This emphasis on customization was intentional. As the Beowulf team noted, ``no two Beowulf Pile-of-PCs\ldots are exactly the same, although they all run the same software'' \cite{ridge1997}. This variability enabled what they termed ``user-driven decisions'' about system evolution, allowing organizations to ``pick and choose from a wide array of sources, try things out, and change the configuration over time'' rather than being ``limited to the vendor's current options lists which may be months out of date'' \cite{ridge1997}. For example, the Computational Biophysics Section at the National Institutes of Health built a Beowulf cluster in 1997 for cost-effective molecular modeling \cite{billings_lobos}.

\subsection*{Beowulf's Legacy and Impact}

Beowulf today is recognized as a pioneering COTS Linux cluster system using networked PCs. NASA calls Beowulf ``the foundation of today's high-end computing systems'' and ``a new model for enabling the efficient storage and retrieval of massive datasets and scalable parallel computing'' \cite{keefe2022}. After NASA built the first Beowulf cluster in 1994, others began building and deploying COTS systems running Linux. One of the advantages of Beowulf systems is their flexibility, which was not available in traditional supercomputers. There is no required Beowulf software, meaning researchers can bring their own software codes and tools to the system. Essentially, Beowulf clusters comprise Linux, the software and tools the researcher brings to the project, and an active community that promotes best practices, provides tutorials, and offers to help each other \cite{layton2025}. Beowulf is not solely responsible for the Linux cluster revolution that now dominates in supercomputing, building as it did on previous NOW and Pile-of-PCs approaches, but they did focus attention on the idea of deploying networked high-end workstations as a low-cost method of achieving more speed and performance. That philosophy was a relatively cheap and easy solution for some business enterprises and research scientists, but the sheer performance still lagged behind special-purpose commercial supercomputers. There was the space for a new project building on the ideas of Linux clusters, with pure performance as the major criterion.

\section{The Roadrunner Approach to Linux Supercomputing}

In 1998 I developed the Roadrunner supercomputer that represented a different emphasis on Linux-based high-performance computing. Roadrunner's architecture integrated commercial off-the-shelf components with specialized high-performance networking technology, specifically incorporating a COTS network technology, Myrinet, which provided low-latency, high-bandwidth network. Unlike the Beowulf project, which was developed by a team of five engineers---Thomas Sterling, Don Becker, John Dorband, Jim Fischer, and Dan Jacob---with NASA institutional support, I designed, built, and deployed Roadrunner entirely on my own. There was no team. As a newly-arrived assistant professor at the University of New Mexico, I served simultaneously as system architect, kernel programmer, network engineer, system administrator, application porter, and benchmarker. Every line of kernel code I modified, every cable I connected, every driver I debugged, and every application I ported to Linux was my own work. This was not a collaborative research project with distributed responsibilities; it was a singular effort to prove that a Linux-based system could compete with commercial supercomputers.

My experience with commodity-based parallel computing predated the Roadrunner project by several years. In 1993, prior to the Beowulf project, I built a parallel computer using Ethernet-connected, Intel-based PCs running the FreeBSD operating system. After completing my Ph.D.\ in May 1996, I spent the next 18 months as a postdoc and National Science Foundation research associate at the University of Maryland's Institute for Advanced Computer Studies (UMIACS). During this period, I constructed an experimental computing cluster comprising 10 DEC AlphaServer nodes, each with four DEC Alpha RISC processors and DEC PCI cards connected to a DEC Gigaswitch ATM switch.

\begin{figure}[!t]
\centering
\includegraphics[width=3.0in]{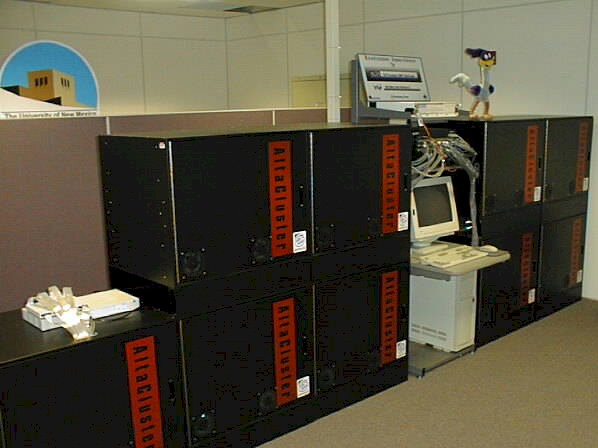}
\caption{Linux prototype on lower-left, and Roadrunner (right). A Myricom dual 8-port SAN Myrinet switch sits on top of the left-most cabinet of the prototype, and four octal 8-port SAN Myrinet switches (not visible) connect Roadrunner. Above Roadrunner's console is a 72-port Foundry Fast Ethernet switch with Gigabit uplinks to the vBNS and Internet. (Image credit: Courtesy of the author.)}
\label{fig:roadrunner_lab}
\end{figure}

My work on Roadrunner began in January 1998 when I joined the University of New Mexico (UNM) as an assistant professor of electrical and computer engineering. As the principal investigator for the UNM Albuquerque High Performance Computing Center (AHPCC) SMP Cluster Computing Project, and working on my own, I constructed the first operational Intel/Linux supercomputer prototype (Roadrunner Phase~1) by April 1998, using eight dual, 333 MHz, Intel Pentium II nodes. This system used Red Hat Linux 5.0b with a custom 2.0.34 SMP kernel, later upgraded to custom 2.1.126 SMP kernel. In addition to the kernel modifications, this prototype required significant engineering work that I undertook single-handedly. Working alone, I ported essential software components to Linux to provide necessary system functionality, modified the Linux kernel and shell for running parallel applications, and ported scientific application codes from National Computational Science Alliance (NCSA) members to Linux---none of which had previously run on the Linux operating system. The Portable Batch System (PBS) job scheduler, originally designed by MRJ Technology Solutions for NASA's supercomputers, required hand-porting to Linux---work I completed without assistance from the original developers or any collaborators. Each component of the system stack, from low-level kernel modifications to high-level application tuning, passed through my hands alone.

Building a production supercomputer as a solo effort required a breadth of expertise that specialized teams typically distribute across multiple engineers. When the kernel crashed---which happened frequently during early development---there was no operating systems specialist to consult; I debugged it myself, often working late into the night tracing through kernel code to identify race conditions or memory management failures in the SMP implementation. When Myrinet drivers needed modification to work with my custom kernel, there was no network engineer to assign the task; I read the specifications, studied the existing driver code, and made the necessary changes. When scientific applications failed to compile or produced incorrect results on Linux, there was no applications team to investigate; I worked directly with the source code, identifying portability assumptions and compiler incompatibilities, then fixing them one by one. The three-network architecture---separating control, data, and diagnostic functions---emerged from my own experience with system failures and my recognition that production reliability required out-of-band management capabilities. Every architectural decision, every technical trade-off, every line of code reflected a single engineer's judgment.

I was no stranger to the movement toward COTS supercomputing. In 1992, while a doctoral student at the University of Maryland, I won the NASA Graduate Student Research Program (GSRP) Fellowship and began a long relationship with John Dorland and Jim Fischer, two scientists at NASA GSFC who later became part of the Beowulf team. At Maryland, my mentor was Joseph F.\ JaJa, a specialist in parallel algorithms \cite{jaja1992}, data structures, and high-performance scientific computing. This mentorship proved instrumental in my ability to conceptualize and eventually implement the architectural approach that would distinguish Roadrunner from other cluster computing paradigms. My discussions and technical exchanges with my NASA colleagues influenced the thinking that would later inform both the Beowulf and Roadrunner projects---though they would ultimately take divergent paths.

An early example of my advocacy for commodity-based parallel computing systems dates to 19 October 1993, when I published a post on the Usenet group comp.parallel suggesting standardized parallel architectures with software abstractions as an alternative to the then-dominant focus on hardware. I also engaged in significant technical discussions with Sterling in Fall 1993, then a University of Maryland instructor and later the main spokesperson for the Beowulf project. I discussed with Sterling my ideas of constructing supercomputers from commodity components and high-performance interconnection networks, unconventional at the time. A commercial deal made a difference in my approach: partnering with Myricom's president and CEO Chuck Seitz to incorporate the first Myrinet interconnection network for Intel/Linux systems. This was a COTS networking technology developed specifically for high-performance cluster computing that provided substantially better performance than the commodity Ethernet used in Beowulf clusters.

A key innovation in Roadrunner was its sophisticated three-network architecture (control, data, and diagnostics), which formed the core of its design philosophy and provided the foundation for its performance and reliability advantages:

\begin{enumerate}
\item A control network (Fast Ethernet with Gigabit Ethernet uplinks).
\item A highly scalable data network (Myrinet switches) for high-bandwidth, low-latency communication.
\item A diagnostic network (chained RS-232 serial ports) to monitor the nodes for failures, provide staged boot up of systems, and enable remote power cycling capabilities for system maintenance.
\end{enumerate}

With its full-duplex 1.28 Gbps bandwidth, Myrinet provided a total bidirectional throughput 256 times that of Beowulf's half-duplex Ethernet, with much lower latency in the tens of microseconds.

\begin{figure}[!t]
\centering
\includegraphics[width=3.0in]{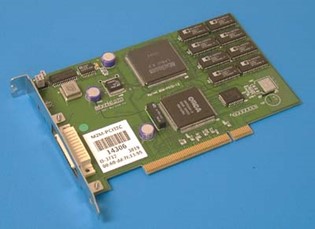}
\caption{Myricom M2M-PCI32c network interface card. (Image credit: CSPi)}
\label{fig:myrinet_nic}
\end{figure}

Roadrunner was the first system to incorporate a high-performance commercial off-the-shelf network into a Linux-based supercomputer. This approach marked a shift from previous supercomputers that relied on proprietary, non-COTS networking solutions, and was different from Beowulf clusters, which used Ethernet. While Myrinet itself was COTS, Roadrunner's networking approach could incorporate any commercially-available networking technology up to the task.

\begin{figure}[!t]
\centering
\includegraphics[width=3.5in]{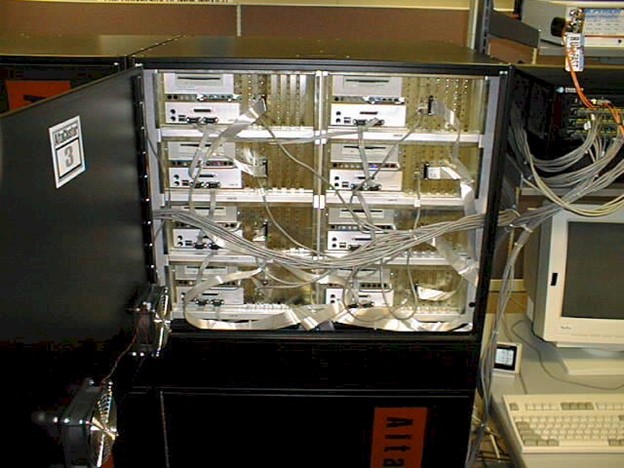}
\caption{Inside a Roadrunner cabinet with each node attached to three networks: Myrinet (ribbon cable), Fast Ethernet (CAT5), and Diagnostic (RS232 serial port). (Image credit: Courtesy of the author.)}
\label{fig:roadrunner_inside}
\end{figure}

Roadrunner represented a strategic architectural midpoint between traditional tightly-coupled supercomputers like Crays and purely network-based parallel systems like Beowulf clusters. While Cray systems achieved parallelism through specialized hardware within a single massive unit, and Beowulf clusters relied entirely on network communication between independent uniprocessor nodes, Roadrunner employed a hybrid approach using dual Intel Pentium II processors within each compute node. This design created a two-tier parallelism model: fine-grained parallelism within each node through shared memory between the dual processors, and coarse-grained parallelism across nodes through the network interconnect. This balanced approach provided greater computational density than pure Beowulf designs while maintaining the cost advantages of commodity hardware, effectively reducing networking overhead for a given level of computational power by performing more work locally within each node before requiring inter-node communication. The multiprocessor node design anticipated the industry's trajectory toward multi-core computing and demonstrated how commodity-based systems could evolve beyond the purely distributed model to incorporate hierarchical parallelism.

NCSA strategically selected a diverse portfolio of national science community supercomputing applications to benchmark Roadrunner Phase~1's performance across multiple computational domains, using these results to inform their decision on proceeding with Roadrunner Phase~2 \cite{bader1999alliance}. The comprehensive application suite included:

\begin{itemize}
\item \textbf{CACTUS}---A modular, manageable high-performance 3D Numerical Relativity Toolkit for solving Einstein equations numerically in computational astrophysics, testing the system's ability to handle gravitational wave physics calculations---a field that would later achieve breakthrough recognition with LIGO's 2017 Nobel Prize-winning detection of gravitational waves.
\item \textbf{MILC}---MIMD Lattice Computation code for quantum chromodynamics (QCD), with the conjugate gradient algorithm for Kogut-Susskind quarks achieving over 60 MFLOPS/node for larger problems.
\item \textbf{ARPI3D}---A 3-D numerical weather prediction model demonstrating the system's capabilities for atmospheric science applications.
\item \textbf{BEAVIS}---Dynamic simulation of particle-laden viscous suspensions, evaluating performance for complex three-dimensional multiphase flow analysis spanning industrial and biological applications.
\item \textbf{AIPS++}---Astronomical Information Processing System for radio astronomy data processing.
\item \textbf{ASPCG}---A 2-D Navier-Stokes solver for computational fluid dynamics.
\end{itemize}

This benchmark suite was carefully chosen to test the system's versatility and computational capabilities across fundamental problems in scientific and engineering contexts, from quantum physics to weather prediction to astrophysics.

Based on demonstrations of this 16-processor prototype, the NSF and NCSA, led then by Larry Smarr, allocated \$400,000 to development. The resulting system, Roadrunner (Phase~2), was fully operational in April 1999 with hardware comprising 64 dual 450 MHz Intel Pentium II processors (128 processors total), 512 KB cache, 512 MB SDRAM with ECC, 6.4 GB IDE hard drives, and Myrinet interface cards. Roadrunner ranked among the 100 fastest supercomputers in the world when it went online in April 1999 \cite{fleck1999}.

The Roadrunner (Phase~2) system development proceeded on the following timeline in 1999:

\begin{itemize}
\item March 17: Received the first 32 nodes
\item March 31: Received the next 32 nodes
\item April 8: Running benchmarks on 64 nodes with Fast Ethernet
\item May 3: Running Alliance test applications and benchmarks on 64 nodes with Myrinet
\item June 1: Began test production
\item June 15: Began test production with Alliance Grid (Globus)
\item July 1: Allocations started for general user community
\end{itemize}

By the time allocations opened in July 1999, Roadrunner had attracted an early adopter user community of 78 researchers from 27 institutions including University of New Mexico (27 users), Rice University (5 users), NCSA (4 users), National Radio Astronomy Observatory (4 users), plus users from University of Texas at Austin, University of Oklahoma, University of Washington, Brown University, University of Illinois, Sandia National Laboratories, Arizona State University, Indiana University, Iowa State University, Los Alamos National Lab, MIT, Pennsylvania State University, Princeton University, University of Kentucky, and New Mexico State University. User projects spanned diverse scientific domains including modelling adhesion/adhesive wear, irregular parallel computation in linear scaling quantum chemistry, PTreeSPH benchmarking, multiphase flow simulation, F15 turbulence simulation, gravitational waves from black hole collisions, molecular dynamics, galaxy formation, parallel least squares finite element methods, particle physics simulations, and weather modeling.

\begin{figure}[!t]
\centering
\includegraphics[width=3.0in]{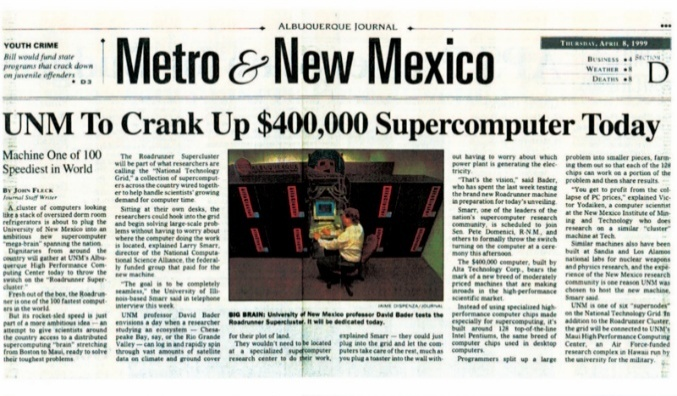}
\caption{The launch of Roadrunner makes the news. ``Machine One of 100 Speediest in World'' with David Bader pictured at Roadrunner's console. (Copyright: The Albuquerque Journal. Reprinted with permission. Permission does not imply endorsement.) \cite{fleck1999}.}
\label{fig:roadrunner_news}
\end{figure}

The system also incorporated specific supercomputing services that were absent in Beowulf clusters, including resource allocation, job scheduling, and monitoring capabilities. Roadrunner's system software included the Red Hat Linux 5.2 operating system with a custom 2.2.10 SMP kernel, sets of compilers from both the GNU Compiler Collection and the Portland Group, and the Portable Batch System (PBS) job scheduler hand-ported to Linux. These management capabilities enabled Roadrunner to function closer to a turnkey commercial capability supercomputing platform that could support multiple simultaneous users running diverse applications across different scientific domains.

\begin{figure}[!t]
\centering
\includegraphics[width=3.5in]{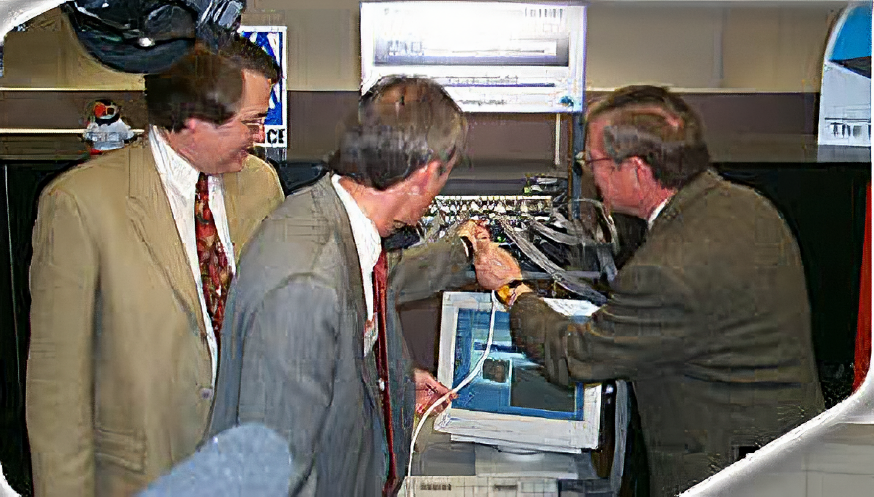}
\caption{NCSA Director Larry Smarr (left), UNM President William Gordon, and U.S.\ Sen.\ Pete Domenici turn on the Roadrunner supercomputer in April 1999. (Image credit: Reprinted with permission of NCSA.)}
\label{fig:roadrunner_launch}
\end{figure}

Unlike Beowulf clusters, which were designed primarily as parallel workstations for local computational tasks, Roadrunner was architected from its inception to be grid-enabled, specifically designed to integrate with distributed scientific infrastructure. Roadrunner became a foundational node on the National Technology Grid, providing researchers across disciplines and across the nation with seamless access to not only supercomputing capabilities but also geographically-distributed scientific instruments, datasets, and other supercomputers from their desktops. The Grid was envisioned as a way to give researchers access to an interconnected ecosystem of computational and scientific resources for large-scale problem solving from their desktops, no matter their location, through the nation's fastest high-performance research networks. Alliance Director Larry Smarr likened the National Technology Grid to the power grid, where users could plug in and get the compute resources they needed, without having to worry about where those resources came from or their own location.

While Beowulf clusters could theoretically be retrofitted with grid middleware and connected to the grid, Roadrunner was purpose-built with grid integration as a core design requirement. This included native support for grid protocols, security frameworks, resource management systems, and the specialized software stack needed to seamlessly share resources across institutional boundaries---capabilities that would require significant additional engineering to retrofit onto Beowulf systems designed primarily for local parallel processing.

\subsection*{A Tool for Enabling Science}

Roadrunner's performance on the Cactus application benchmark showed near-perfect scalability, outperforming systems such as NASA's Beowulf cluster, NCSA's Microsoft Windows NT cluster \cite{hpcwire1998nt}, and Silicon Graphics' Origin 2000 \cite{bader1999chautauqua}. Roadrunner's design philosophy---optimizing for performance rather than simply minimizing cost---produced a system that could effectively compete with traditional proprietary supercomputers on both performance and price. Roadrunner's entry into production in April 1999 as part of the National Science Foundation's National Technology Grid gave researchers across disciplines access to supercomputing capabilities from their desktops and established a blueprint for Linux supercomputing that would eventually become the dominant architecture in high-performance computing.

Edward Seidel, former head of the Numerical Relativity and E-Science Research Groups at the Albert Einstein Institute and now President of the University of Wyoming, recalls: ``It was a very exciting time; Linux clusters were emerging as a huge force to democratize supercomputing and software frameworks providing community toolkits to solve broad classes of science and engineering problems were also taking shape. The collaboration we had between the Cactus team at the Albert Einstein Institute in Germany and David Bader's team with the Roadrunner supercluster was a pioneering effort that helped these movements gain traction around the world. The collaboration helped advance the goals of the Cactus team, led by Gabrielle Allen, whose efforts continue to this day as the underlying framework of the Einstein Toolkit. That toolkit now powers many efforts globally to address complex problems in multi-messenger astrophysics'' \cite{seidel2021}.

Roadrunner transformed computational physics and astronomy research across multiple domains. Astrophysics researchers utilized the system's dual-processor architecture and Myrinet networking for large-scale cosmological simulations, with teams like Brandon Allgood's from UC Santa Cruz running PKDGRAV N-body codes to simulate 1.3 million particles representing cosmic structure formation \cite{allgood1999}. Climate scientists, including Dan Weber and Kelvin Droegemeier, leveraged Roadrunner's superior network performance for weather prediction codes and detailed thunderstorm simulations, finding significant improvements in forecast turnaround time and resolution \cite{bader2021linux}. In fundamental physics, Steven Gottlieb's lattice quantum chromodynamics research used Roadrunner's balanced architecture for MILC collaboration calculations studying quark-gluon interactions, achieving sustained performance of 1.25--2.0 Gflops and enabling breakthrough observations of meson decay on the lattice \cite{bernard2001, christ2000, gottlieb2001, gottlieb2001b}. The system also enabled quantum chemistry advances through linear scaling SCF calculations and large-scale electronic structure studies of molecular systems like water clusters and polymers \cite{challacombe2000}.

Engineering and materials science applications demonstrated Roadrunner's versatility across computational domains. In fluid dynamics, researchers used the system for parallelized RMA2 hydrodynamic modeling \cite{rao2005}, advanced aircraft simulations using spectral/hp element methods for F-15 configurations \cite{karniadakis2000}, breakthrough parallel multipole algorithms for vorticity-based CFD methods that reduced computational complexity from $O(N^2)$ to $O(N \log N)$ \cite{brown2003}, and environmental flow studies of toxic chemical dispersion with up to 29 million unknowns \cite{tsang1999}. Computational electromagnetics researchers performed the first full-scale numerical simulations of Maxwell's equations for ultrashort optical pulses in nonlinear media \cite{bennett2000, bennett2003}, while materials scientists conducted Monte Carlo simulations of charge carrier transport in organic semiconductors using $64^3$ lattice sites \cite{novikov2003a, novikov2004, novikov2001}. Advanced materials research included comprehensive first-principles quantum mechanical calculations of defect physics and radiation effects in silicon dioxide for semiconductor applications \cite{karna2000, pineda2000, pineda2001}.

Beyond traditional scientific computing, Roadrunner pioneered parallel algorithm development and distributed computing research. My research group developed novel graph algorithms achieving nearly linear speedup on problems with up to 256 million vertices \cite{bader1999cycle, bader2000hpc, bader2000ear}, while Mohammad Mikki's distributed Barnes-Hut tree codes demonstrated 10--45\% performance improvements through optimization techniques tested with up to 64,000 particles \cite{mikki2002}. The system also served as a testbed for hierarchical broadcast algorithms showing 20--30\% improvements over standard MPI implementations \cite{sun2002}, parallel application sensitivity measurement tools \cite{leon2003diss, leon2003mpi}, distributed software design models \cite{dwivedula2002}, and the Adaptive Distributed Virtual Computing Environment (ADViCE) for middleware and virtualization research \cite{kim2000}. This diverse research portfolio established Roadrunner as a transformative platform that democratized supercomputing access and validated Linux-based clusters as viable alternatives to expensive proprietary systems across the computational science spectrum.

\subsection*{Impact and Legacy of Roadrunner}

The University of New Mexico's Albuquerque High Performance Computing Center pioneered a collaborative model between academic institutions and commercial Linux vendors that would reshape the supercomputing industry. An early example of this partnership was ``BlackBear,'' a 16-node Linux cluster built around dual Intel Pentium III 550 MHz processors per node with 512 MB of SDRAM, running VA Linux Systems' Red Hat 6.0.4 distribution with a Linux 2.2.12smp kernel. BlackBear featured a dual-network architecture combining 10/100 BaseT Ethernet for control functions with a Myrinet interconnect for high-speed data communication, while its name honored New Mexico's official state animal and cultural heritage. This technical partnership with VA Linux Systems demonstrated the growing commercial viability of Linux-based supercomputing solutions and established a template for future academic-industry collaborations.

Building on this collaborative foundation, I also led the development of ``LosLobos'' with IBM following Roadrunner's construction---IBM's first Linux production system and a significant escalation in scale and ambition \cite{unm2000loslobos}. LosLobos premiered at number 24 on the Top500 supercomputer list in summer 2000, featuring 256 dual-processor Intel-based IBM servers with Myrinet connections (512 processors total) capable of 375 gigaflops, running Red Hat Linux 6.1 with a custom 2.2.13smp kernel. The knowledge IBM gained through this UNM collaboration enabled the company to create its first preassembled and preconfigured Linux server clusters for business within just one year, marking a crucial transition from experimental academic systems to commercial products.

These successful demonstrations of Linux-based high-performance computing attracted widespread industry attention and investment. Companies including IBM, VA Linux Systems, and Apple, established direct relationships with UNM to access the hardware configurations, software implementations, and design methodologies that had proven Linux clusters could compete with traditional supercomputers. The commercial stakes became evident when VA Linux Systems capitalized on this growing enthusiasm by going public on December 9, 1999, with a record-breaking IPO that saw its stock price surge 737\% on the first day of trading \cite{glasner1999}, reflecting both investor confidence in Linux's potential and broader market recognition of the paradigm shift toward open-source supercomputing. These academic-industry partnerships created crucial bridges that rapidly translated technical breakthroughs into market-ready products, democratizing access to supercomputing capabilities across the industry.

Roadrunner's architectural innovations and operational success established a template for modern Linux-based supercomputing that continues to influence high-performance computing design and deployment strategies across the global research enterprise. This was solidified through Roadrunner's performance achievements and practical viability. By successfully executing real-world scientific applications across multiple domains, Linux-based systems like Roadrunner provided concrete evidence that commodity-based architectures could serve as viable alternatives to traditional proprietary supercomputers in production environments. This proof of concept was particularly significant because it addressed longstanding concerns about the reliability, performance, and scalability of open-source computing platforms for mission-critical scientific research. The validation provided by Roadrunner's success accelerated industry adoption: within just a few years, the approach developed by Roadrunner became the dominant paradigm in supercomputing, fundamentally altering market dynamics and procurement strategies.

Larry Smarr, Founding Director of NCSA, recalled the historical importance of Roadrunner's development: ``One of the most significant events that occurred in this period was when David [Bader] at University of New Mexico as a member of the Alliance created the first commercial off-the-shelf supercomputer, in other words a supercomputer built of PC server technologies, and he put it on the National Technology Grid. So here was a commodity-built, PC-based endpoint going into the technology grid\ldots\ This is an historic event. It took resources from the Alliance, but it took David's creative energies and innovation to do that'' \cite{smarr2021}.

Roadrunner's achievement validated that democratized computing approaches could scale to compete with the world's most powerful systems, representing perhaps the most profound societal impact of its architectural approach. Prior to the emergence of Linux-based supercomputing, access to high-performance computing capabilities was largely restricted to well-funded government laboratories, major research universities, and large industrial corporations that could afford the substantial capital investments required for proprietary systems. The cost barriers were not merely financial but also technical, as these systems required specialized expertise for operation and maintenance that was scarce and expensive. Roadrunner's demonstration that supercomputing capabilities could be achieved through more accessible technologies fundamentally altered this equation. Linux-based supercomputing opened pathways for a broader range of institutions to participate in cutting-edge computational research and innovation. Smaller universities, regional research institutions, and emerging technology companies could now access computational capabilities that had previously been beyond their reach. This democratization fostered a more inclusive scientific and technological ecosystem, enabling research breakthroughs from previously underrepresented institutions and geographic regions. The ripple effects extended beyond traditional academic and industrial research settings, as the reduced barriers to entry allowed innovative applications of supercomputing to emerge in fields ranging from financial modeling to entertainment and media production.

The economic impact of my Linux-based supercomputing design has been transformative on a global scale. Hyperion Research \cite{joseph2022} quantified the remarkable economic contribution of this technological shift, finding that over the 25 years following Roadrunner's introduction, Linux-based HPC systems contributed to the development of products and services worth more than \$100 trillion globally. This staggering figure reflects not only direct economic activity but also the multiplier effects of scientific discoveries, technological innovations, and industrial breakthroughs enabled by accessible high-performance computing. The economic impact extends across virtually every sector of the economy, from pharmaceutical development and materials science to climate modeling and artificial intelligence research. Most recently, the critical role of Linux-based HPC became evident during the COVID-19 pandemic, when these systems powered the computational research that enabled rapid vaccine development, epidemiological modeling, and public health response strategies \cite{brase2022}. The ability to rapidly deploy computational resources for urgent societal challenges demonstrated the strategic importance of maintaining robust, accessible supercomputing infrastructure based on open technologies.

As the global economy continues to evolve and worldwide challenges increasingly threaten human wellbeing---from climate change and pandemics to resource scarcity and geopolitical instability---Linux supercomputers continue to serve as the powerhouse systems that drive economic growth, solve complex problems, and ensure collective safety and security. The architectural template established by Roadrunner has proven remarkably durable and adaptable, providing a foundation for continuous innovation in computational science and engineering that remains as relevant today as it was at the dawn of the new millennium.

\section{Conclusions}

The divergent approaches represented by Beowulf and Roadrunner provide insights into high-performance system design philosophy. Beowulf clusters emerged as an accessible approach to parallel computing, democratizing access through a design philosophy focused exclusively on mass-market components. The Beowulf project brought awareness to the potential of COTS components in HPC; however, this approach came with certain architectural constraints, particularly in network performance and system management capabilities, which affected its suitability for communication-intensive workloads and multi-user environments.

Roadrunner's design philosophy took a different path, integrating commercial off-the-shelf components with specialized networking technology and implementing comprehensive resource management capabilities. This balanced approach prioritized overall system performance and usability, enabling the support of multiple simultaneous users across diverse scientific domains. By optimizing for computational efficiency and flexibility, Roadrunner established a blueprint for Linux supercomputing that effectively combined the cost advantages of commodity components with the performance characteristics needed for demanding scientific applications.

The fact that Roadrunner was designed, built, and deployed by a single individual---without the institutional infrastructure, dedicated staff, or team-based approach that characterized both the Beowulf project and traditional supercomputer development---demonstrates that transformative contributions to high-performance computing need not require large teams or extensive resources. What Roadrunner required was a willingness to work across the entire system stack, from kernel internals to application algorithms, and the determination to solve every problem personally rather than delegating to specialists who did not exist.

This historical comparison illuminates an important lesson for computing system design: system developers can choose different approaches based on their target applications and user communities. The Beowulf team's strict adherence to mass-market commodity components successfully served individual researchers seeking affordable parallel computing access, while Roadrunner's integration of commodity and specialized components addressed the needs of multi-user supercomputing environments. Both approaches made valuable contributions to the evolution of Linux-based high-performance computing. Beowulf demonstrated the viability of commodity cluster computing and democratized access to parallel processing, while Roadrunner established an architectural template that balanced cost-effectiveness with performance requirements. The success of both approaches demonstrates that effective system design depends on clearly understanding target requirements and user needs rather than adhering to a single universal philosophy.

The validation of Roadrunner's architectural approach is demonstrated by examining the specific design elements that define modern supercomputing. By 2017, Linux-based systems achieved 100\% dominance of the world's fastest 500 supercomputers \cite{linuxfoundation2017}, but more significantly, these systems universally adopted Roadrunner's core architectural principles rather than Beowulf's design philosophy. Modern supercomputers employ multi-core nodes connected via high-performance interconnects---exactly the template Roadrunner established with its dual-processor nodes and Myrinet networking, not Beowulf's uniprocessor-plus-Ethernet approach. Today's exascale computing systems, capable of calculating at least $10^{18}$ floating point operations per second, require the sophisticated resource management, multi-user capabilities, and scalable networking architecture that Roadrunner pioneered. While Beowulf demonstrated the viability of commodity components, its design constraints---single processors per node, commodity Ethernet networking, and lack of system management---proved inadequate for the communication-intensive, massively parallel workloads that define serious supercomputing. Roadrunner's architectural template succeeded because it combined commodity economics with performance-oriented design choices, creating a blueprint that could scale from hundreds to millions of processors while maintaining the cost advantages that made Linux supercomputing accessible.

The architectural template that Roadrunner established---commodity processors, high-performance interconnects, multi-user resource management---now underpins systems from departmental clusters to exascale machines. The lessons from this transition period remain instructive: that effective parallel system design requires matching architectural choices to target applications and user communities, and that open platforms can achieve performance competitive with proprietary alternatives. As the field confronts new challenges in heterogeneous computing, AI acceleration, and energy-efficient design, the debates between accessibility and performance that distinguished Beowulf from Roadrunner will continue to shape the next generation of parallel and distributed systems.

\section*{Acknowledgments}

The work of David A.\ Bader is supported in part by NSF grants CCF-2109988, OAC-2402560, and CCF-2453324. The author wishes to thank Troy Kaighin Astarte, Lecturer at Swansea University, and Brian Carpenter, honorary professor at the University of Auckland, for their contributions to this article. We acknowledge Karen Green for her copyediting that improved the readability of this article.

\vspace{1em}

\noindent\textbf{David A.\ Bader} is a Distinguished Professor and founder of the Department of Data Science and inaugural Director of the Institute for Data Science at New Jersey Institute of Technology. Prior to this, he served as founding Professor and Chair of the School of Computational Science and Engineering, College of Computing, at Georgia Institute of Technology. Bader is a Fellow of the IEEE, ACM, AAAS, and SIAM and has co-authored more than 400 scholarly papers. Bader is the 2021 recipient of the IEEE Sidney Fernbach Award; the 2022 Innovation Hall of Fame inductee of the University of Maryland's A.\ James Clark School of Engineering; a 2025 inductee of the Mimms Museum of Technology and Art's Hall of Fame; and the 2025 recipient of the Heatherington Award for Technological Innovation. In 2025, \emph{HPCwire} named Bader as one of its ``35 Legends'' of HPC.

\end{document}